\documentclass[reprint,amsmath,amssymb,aps,floatfix,showpacs,longbibliography,
prl,
]{revtex4-1}

\usepackage[utf8]{inputenc} 
\usepackage[babel]{csquotes} 

\usepackage{graphicx}
\usepackage{dcolumn}
\usepackage{bm}
\usepackage[]{units}	
\usepackage{xcolor}
\usepackage{soul}
\usepackage{changes}
\usepackage{hyperref}
\usepackage[all]{hypcap}
\usepackage{textcomp}
\usepackage{upgreek}
\usepackage{enumerate}
\usepackage{multirow} 

\newcommand{\ket}[1]{\ensuremath{\left|#1\right\rangle}}
\newcommand{\abs}[1]{\ensuremath{{\left\lvert {#1} \right\rvert}}}

\newcommand{\reffig}[1]{Fig.~\ref{#1}}

\begin{document}

\title{High-resolution spectroscopy on cold electrically trapped formaldehyde}

\author{Alexander Prehn}
\author{Martin Ibr\"ugger}
\author{Gerhard Rempe}
\author{Martin Zeppenfeld}
	\email{Martin.Zeppenfeld@mpq.mpg.de}
\affiliation{Max-Planck-Institut f\"ur Quantenoptik, Hans-Kopfermann-Strasse 1, 85748 Garching, Germany}

\date{\today}

\begin{abstract} 
	We present precision spectroscopy on electrically trapped formaldehyde (H$_2$CO), demonstrating key attributes which will enable molecular spectroscopy with unprecedented precision. Our method makes use of a microstructured electric trap with homogeneous fields in the trap center and rotational transitions with minimal Stark broadening at a 'magic' offset electric field. Using molecules cooled to the low millikelvin temperature regime via optoelectrical Sisyphus cooling, we reduce Stark broadening on the $J=5\leftarrow4$ ($K=3$) transition at 364\,GHz to well below 1\,kHz, observe Doppler-limited linewidths down to 3.8\,kHz, and determine the line position with sub-kHz uncertainty. Our results and clear prospects for even narrower spectra pave the way towards in-trap precision spectroscopy on diverse molecule species.
\end{abstract}


\maketitle


Precise measurements of atomic transition frequencies, now reaching relative accuracies below $10^{-17}$~\cite{Rosenband2008}, have enabled groundbreaking progress in science and technology, ranging from optical clocks~\cite{Ludlow2015} to tests of fundamental physical theories~\cite{Griffith2009}. Despite the lower accuracy achieved with spectroscopic investigations of molecules so far, the structure and symmetry of molecular systems can often provide a more sensitive probe of fundamental physics~\cite{Steimle2014,DeMille2017}. A prominent example is the search for a permanent electric dipole moment of the electron~\cite{Baron2014,Cairncross2017}. Molecules are also used to test parity violation~\cite{Quack2011} and to measure fundamental constants~\cite{Mejri2015,Biesheuvel2016} and their possible time variation~\cite{Shelkovnikov2008,Truppe2013}. Moreover, precise knowledge of molecular constants is required in cold chemistry and collision studies~\cite{Bell2009,Bohn2017} as well as for the interpretation of astrophysical spectra, a research area where formaldehyde (H$_2$CO) is of immense importance~\cite{Bergin2007,Muller2017}.

Such questions are addressed with high-resolution spectroscopy on structurally and chemically diverse molecule species. So far, most precision experiments with neutral molecules use beams of cold particles, limiting the interrogation time to a few milliseconds~\cite{Veldhoven2004,Hudson2006a,Truppe2013,Baron2014,DeNijs2014,Cahn2014}. 
Longer interaction times can be achieved by launching molecules in a fountain~\cite{Bethlem2008,Tarbutt2013} which was demonstrated with ammonia ($^{14}$NH$_3$) molecules~\cite{Cheng2016} but not yet used for spectroscopy. For very long interaction times and highest sensitivity, however, high-precision experiments with directly cooled and trapped molecules are desired~\cite{Wall2016,Kozyryev2017}. 

Apart from investigations with molecular ions~\cite{Biesheuvel2016,Cairncross2017,Alighanbari2018}, precision spectroscopy on trapped molecules has only been performed with diatomic species associated from ultracold atoms~\cite{McGuyer2015,Park2017} for two reasons: First, direct cooling techniques reached the ultracold ($T<1\,\mathrm{mK}$) temperature regime only very recently~\cite{Prehn2016,Norrgard2016}. Second, the possibly very long interrogation time accessible with trapped molecules comes at the expense of Stark or Zeeman shifts of the molecular energy levels, usually broadening the spectral lines under investigation.

Here, we solve the latter problem by transferring concepts known from atomic clocks~\cite{Ye2008} to cooled formaldehyde molecules held in our unique boxlike electrostatic trap featuring a tunable homogeneous offset field~\cite{Englert2011}. Specifically, we perform microwave (MW) spectroscopy on a rotational \enquote*{clock} transition with vanishing first-order differential Stark shift and measure at a \enquote*{magic} offset electric field with minimal and currently negligible residual Stark broadening. Using optoelectrical Sisyphus cooling we prepare about $10^6$ molecules at a temperature of a few mK and in a single rotational state~\cite{Prehn2016}. Thus, we observe Doppler-limited linewidths down to 3.8\,kHz and are able to determine the center frequency with sub-kHz uncertainty. 
We model the spectral lineshape which is influenced by two parameters, the kinetic energy of the molecules and the offset field in the trap. By systematically varying both of them we verify our model and demonstrate a good theoretical understanding of the performed experiments. 
With the observed precision we do not surpass conventional techniques employing molecular beams yet. Nevertheless, to the best of our knowledge, this proof-of-principle experiment presents the narrowest linewidths observed with electrically trapped neutral molecules so far.

\begin{figure}[b]
	\centering
	\includegraphics{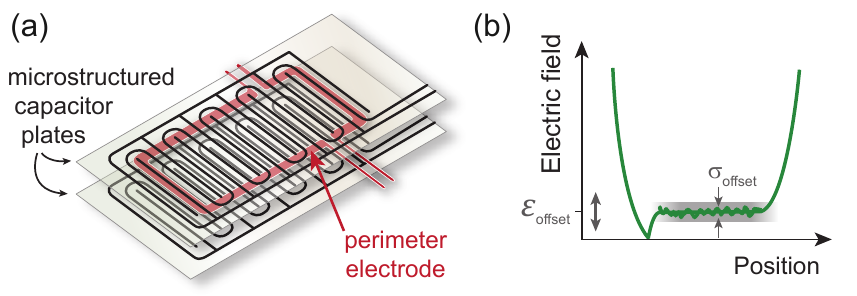}
	\caption{(color online)
		(a)~Illustration of the microstructured electric trap (not to scale). 
		(b)~Simplified sketch of the position-dependent electric fields in the trap. The main features are high trapping fields at the edges, a tunable homogeneous offset field $\mathcal{E}_\mathrm{offset}$ with a finite roughness $\sigma_\mathrm{offset}$, and isolated regions of low electric field with $\mathcal{E}<\mathcal{E}_\mathrm{offset}$ which will play a role in the final part of this paper.
	}
	\label{fig:trap}
\end{figure}

\begin{figure*}[tb]
	\centering
	\includegraphics{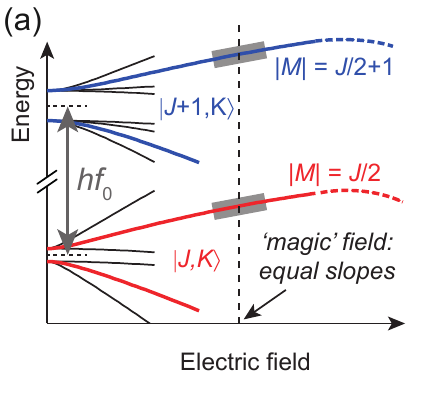}
	\includegraphics{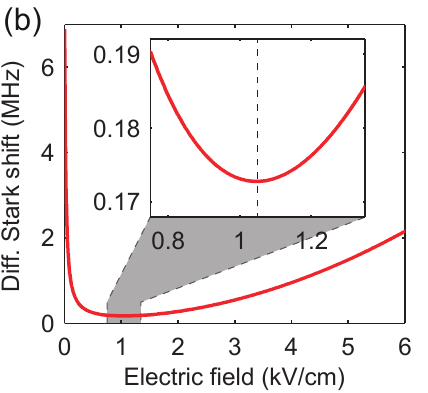}
	\includegraphics{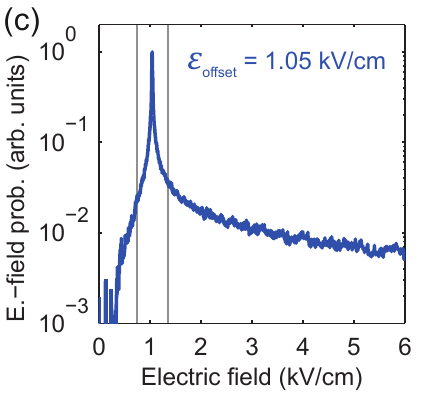}
	\includegraphics{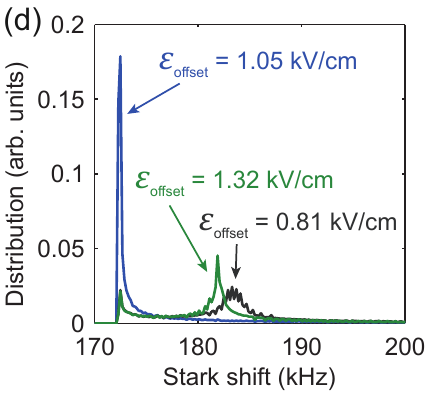}
	\caption{(color online)
		(a)~Sketch of Stark-shifted energy levels of two inversion-doublet rotational states $\ket{J,K}$ and $\ket{J{+}1,K}$ of a slightly asymmetric rotor. For low fields a quadratic Stark shift is observed due to the finite inversion splitting, while the Stark shift is almost perfectly linear for intermediate  fields. For very large field strengths the molecular energy levels bend quadratically because the influence of neighboring $J$-states coupled by the electric field dominates the Stark shift. A \enquote*{clock} transition connecting two states with approximately equal first-order linear Stark shift features a dramatically reduced differential Stark shift. Variation of the differential Stark shift is minimal in the \enquote*{magic} field where the two energy levels have identical slopes. In this paper, all Stark shifts and detunings are specified with respect to the zero-field center frequency $f_0$ coupling two inversion doublets.
		(b)~Differential Stark shift of the transition $\ket{4,3,2}\leftrightarrow\ket{5,3,3}$ vs. electric field. The minimum is marked with a dashed line in the inset. At the \enquote*{magic} field, the initial and final states possess individual Stark shifts of about 370\,MHz. The shift is specified relative to $f_0=364282.0128\,\mathrm{MHz}$ which was calculated from literature data~\cite{Cornet1980}.
		(c)~Simulated electric-field distribution in the trap for the \enquote*{magic} trapping configuration. The gray lines mark the electric-field range of the inset of panel~(b). The resulting offset electric field is specified. The FWHM of the distribution is about 15\,V/cm.
		(d)~Distribution of differential Stark shifts in the trap for the transition $\ket{4,3,2}\leftrightarrow\ket{5,3,3}$ and the \enquote*{magic} trapping configuration as well as two other offset electric fields. The distributions are normalized to equal areas under the curves. With the \enquote*{magic} offset field, Stark broadening is reduced to below 1\,kHz.
	}
	\label{fig:Stark}
\end{figure*}

Experiments are performed in our microstructured electrostatic trap. It consists of two capacitor plates holding an array of parallel electrodes and a perimeter electrode providing confinement from the side, as sketched in \reffig{fig:trap}(a). This arrangement produces strong electric fields at the boundaries of the trapping volume ($\sim50\mathrm{kV/cm}$ corresponding to a nominal trap depth of $\sim1\,\mathrm{K}$), a tunable offset electric field $\mathcal{E}_\mathrm{offset}$ with a finite roughness $\sigma_\mathrm{offset}$ in the center~\cite{Prehn2016}, and results in a three-dimensional boxlike potential for molecules in low-field-seeking states. The main features of the trapping potential are visualized in \reffig{fig:trap}(b). The trap design~\cite{Zeppenfeld2013} as well as its integration into the experimental apparatus and performance have been described earlier~\cite{Englert2011,Gloeckner2015,Prehn2016}. Cooled formaldehyde molecules are stored in the trap with 1/e decay times of up to one minute~\cite{Prehn2016}. 

We exploit the generic structure of the slightly asymmetric rotor formaldehyde for both strong electric trapping and precise spectroscopy with minimal Stark broadening. Due to its slight asymmetry formaldehyde features many pairs of opposite-parity rotational states with a small inversion splitting~\cite{Townes1975}. These inversion doublets are coupled strongly already in static fields much smaller than the offset electric field of our trap, and hence essentially possess the properties of symmetric-top states~\cite{Prehn2016}. In particular, they feature large linear Stark shifts over a wide range of electric fields, thus facilitating trapping. Following the correspondence to symmetric-top states,  we label states with symmetric-top rotational quantum numbers $J$, $K$, and $M$ as $\ket{J,\mp K,\pm M}$ with $\mp K$ chosen positive. 

Stark broadening in a spectroscopy experiment is caused by the \emph{differential} Stark shift, i.e., the difference of the Stark shifts of the initial and final states. To essentially eliminate Stark broadening, the variation of the differential Stark shift as a function of electric field strength is minimized. First, we eliminate the largest contribution, the linear first-order differential Stark shift, by choosing a \enquote*{clock} transition which connects states with equal first-order Stark shifts $\nu_s=-\frac{\mu\mathcal{E}}{h}\frac{KM}{J(J+1)}$~\cite{Townes1975} (with electric dipole moment $\mu=2.33\,\mathrm{Debye}$~\cite{Fabricant1977}, electric-field strength $\mathcal{E}$ and Planck constant $h$). Since molecules should remain trapped during the entire experiment, we desire low-field-seeking initial and final states with equal, non-zero $\nu_s$. Restricting ourselves to purely rotational, electric-dipole allowed transitions, the conditions are fulfilled for transitions $\ket{J,K,M{=}J/2}\leftrightarrow\ket{J{+}1,K,M{+}1}$. Second, we tune the offset electric field of our trap to a \enquote*{magic} value with minimal variation of the residual higher-order differential Stark shift. Both aspects are illustrated and explained in \reffig{fig:Stark}(a).

For further investigations we pick the \enquote*{clock} transition $\ket{4,3,2}\leftrightarrow\ket{5,3,3}$. Its differential Stark shift is shown in \reffig{fig:Stark}(b).  It was calculated by numerically diagonalizing the full rotational Hamiltonian in a static electric field using the rotational constants of Ref.~\cite{Brunken2003}. As expected, the computed shift is comparably small and features a minimum which defines the \enquote*{magic} field for spectroscopy, with small variation even for large changes in electric field. Note that the position of the minimum strongly depends on the specific \enquote*{clock} transition, while the shape of the curve is similar for other such transitions.

The simulated electric-field distribution in our trap corresponding to the \enquote*{magic} trapping configuration with $\mathcal{E}_\mathrm{offset}=1.05\,\mathrm{kV/cm}$ is plotted in \reffig{fig:Stark}(c). The strongly peaked graph with a relative full width at half maximum of only 1.4\,\% shows that the trap potential indeed is boxlike (see also Ref.~\cite{Prehn2016}). The width of the distribution quantifies the aforementioned roughness $\sigma_\mathrm{offset}$ of the floor of the boxlike potential. We verified the simulated distribution experimentally using a previously developed method relying on Stark spectroscopy~\cite{Gloeckner2015,Prehn2016}. In fact, measured distributions are slightly narrower than the simulation for large offset electric fields ($\mathcal{E}_\mathrm{offset}>100\,\mathrm{V/cm}$)~\cite{Prehn2016}.

Combining the differential Stark shift and the field distribution, we calculate the distribution of Stark shifts in the trap for the \enquote*{magic} and two additional trapping configurations. The resulting curves are shown in \reffig{fig:Stark}(d) and represent the expected Stark broadened lineshape of the investigated transition for a specific electric field configuration. Stark broadening is minimized and reduced to below 1\,kHz with the \enquote*{magic} offset electric field. In contrast, a double-peaked shape is found for offset electric fields which are chosen such that the differential Stark shift is about 10\,kHz larger in the respective offset electric field, with one offset field being smaller and one larger than the \enquote*{magic} field. Here, the peaks on the right occur because a range of Stark shifts is enhanced by the detuned offset electric field. Common to all three curves is the sharp cutoff at the left side due to the minimum of the differential Stark shift. In principle, this feature permits a very precise determination of the line position.

\begin{figure}[tb]
	\centering
	\includegraphics{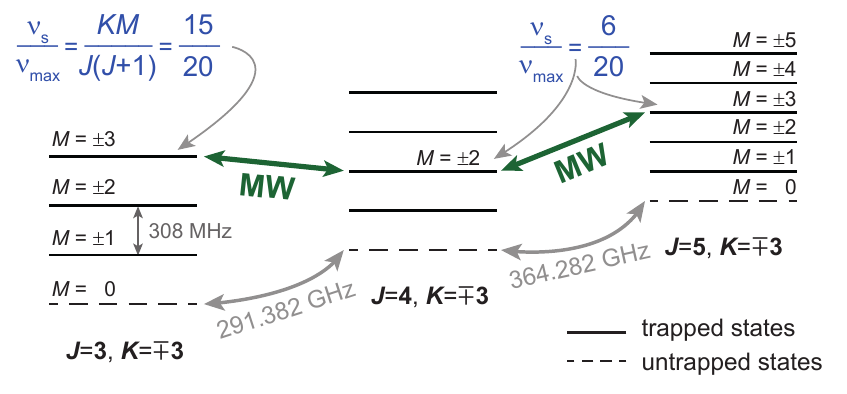}
	\caption{(color online)
		Level scheme for microwave (MW) spectroscopy of the rotational \enquote*{clock} transition $\ket{4,3,2}\leftrightarrow\ket{5,3,3}$ in the ground electronic and vibrational state, with a static electric field~$\mathcal{E}$ applied. Initially, the molecules are prepared in the state $\ket{3,3,3}$. The spectroscopy signal is obtained by coupling the states $\ket{3,3,3}$, $\ket{4,3,2}$, and $\ket{5,3,3}$ with MW as shown and measuring the signal of molecules originating from the states $\ket{5,3,M}$ afterwards. The fractional first-order Stark shifts $\nu_s/\nu_\mathrm{max}=\nu_s/\left(\mu\mathcal{E}/h\right)$ are specified for the relevant states. The separation of $M$ sublevels is given for $J=3$ for the \enquote*{magic} offset electric field.
	}
	\label{fig:scheme}
\end{figure}

For trapping, cooling, and spectroscopy we use the low-field-seeking components of the rotational states characterized by $J=3,4,5$ and $\abs{K}=3$. The resulting level scheme is sketched in \reffig{fig:scheme}. After optoelectrical Sisyphus cooling~\cite{Zeppenfeld2009,Zeppenfeld2012}, the molecules are prepared in a single rotational state via optical pumping~\cite{Glockner2015a}. Since efficient state preparation requires a maximally polarized target state $\ket{J,K,M{=}J}$~\cite{Glockner2015a}, we chose $\ket{3,3,3}$. As a consequence, the ensemble of cooled formaldehyde molecules populates this state with a purity of over 80\,\%~\cite{Prehn2016}. Then, the offset field is adjusted to the \enquote*{magic} value and two MW frequencies are applied to transfer population to the state $\ket{5,3,3}$ via the \enquote*{clock} transition under investigation. Finally, state detection (via MW depletion~\cite{Gloeckner2015,Prehn2016}) is employed to measure the signal of molecules in the states $\ket{5,3,M}$ as a function of the MW frequency driving the \enquote*{clock} transition.

\begin{figure*}[tb]
	\centering
	\includegraphics{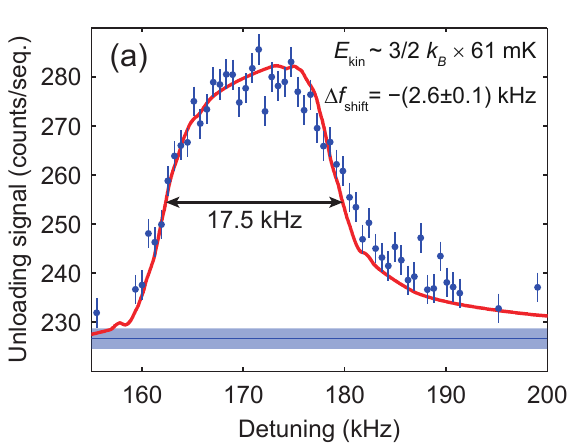}
	\includegraphics{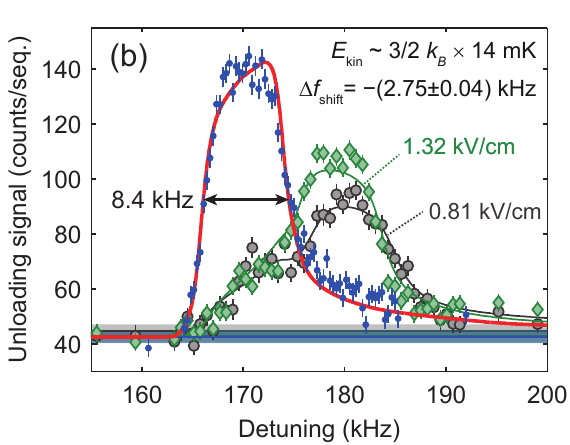}
	\includegraphics{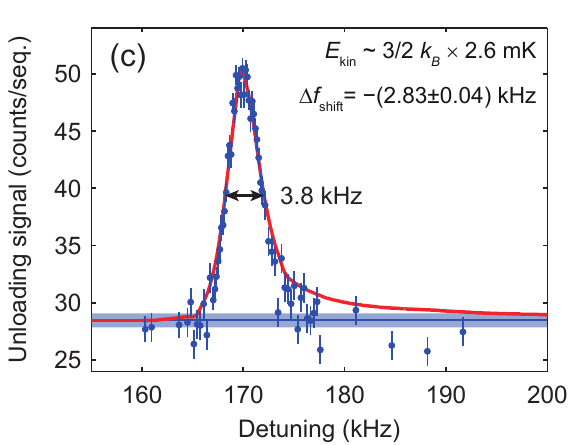}
	\caption{(color online)
		Measured and simulated spectra. (a) -- (c) Data at the \enquote*{magic} offset electric field was recorded for molecule ensembles with varying kinetic energy (median kinetic energy of the ensemble specified in the upper right corner of each panel, cf. Ref.~\cite{Prehn2016}). For the ensemble with intermediate energy the offset electric field was varied~(b). As expected, the spectra obtained with offset electric fields distinct from the \enquote*{magic} configuration (curves labeled with the respective $\mathcal{E}_\mathrm{offset}$) are much broader. The detuning is specified relative to the center frequency 364282.0128\,MHz (see \reffig{fig:Stark}). The vertical scaling and the horizontal position ($\Delta f_\mathrm{shift}$ only given for \enquote*{magic}-field data) of the simulated spectra were determined with a least-$\chi^2$ fit. The horizontal lines mark the signals measured with the MW left off and determine the base lines of the modeled spectra. Error bars and shaded regions specify the 1\,$\sigma$ statistical error.
	}
	\label{fig:alternative}
\end{figure*}

The spectral shape is modeled taking into account only the differential Stark shift, the trap electric fields and Doppler broadening. To this end, we convolve the differential-Stark-shift distributions of \reffig{fig:Stark}(d) with expected Doppler-broadened spectral profiles derived from measured distributions of the molecular kinetic energy. The kinetic energy of the molecules in the \enquote*{magic} trapping configuration was measured with rf knife-edge filters, as explained in detail elsewhere~\cite{Prehn2016}. A spectral profile of Doppler shifts was then numerically calculated assuming an isotropic velocity distribution in the trap.

Measured and simulated spectra of the chosen \enquote*{clock} transition, obtained under different experimental conditions, are presented in \reffig{fig:alternative}. We first focus on the trapping fields. To test our model, we used a molecule ensemble with intermediate kinetic energy to record spectra at different offset electric fields. As shown in \reffig{fig:alternative}(b), the calculated curves explain the observed shape. The good overlap of experiment and theory demonstrates a good control and understanding of the relevant experimental parameters. As expected, the narrowest spectrum is recorded with the \enquote*{magic} offset electric field. Note that the subtle differences observed in the double-peaked distributions of Stark shifts for the remaining two trappping configurations [\reffig{fig:Stark}(d)] are also visible as structure in the measured data.

By reducing the molecules' kinetic energy and measuring at the \enquote*{magic} offset electric field we are able to reduce the linewidth to 3.8\,kHz for the coldest ensemble ($T\sim2.6\,\mathrm{mK}$), corresponding to a relative width of $10^{-8}$ [\reffig{fig:alternative}(c)]. The spectra recorded with varying molecular energy [reduced by a factor $>20$ from \reffig{fig:alternative}(a) to (c)] show that the linewidth is indeed limited by Doppler broadening, whereas the asymmetry of the spectrum is caused by the minimum of the differential Stark shift.

The relatively sharp cutoff at lower frequencies allows for a precise determination of the line position. We distinguish two aspects. The position of the Stark shifted line can be deduced directly from the measurement. From the fit of the theory curve to the data, the position is determined with sub-kHz statistical uncertainty (see fitted parameter $\Delta f_\mathrm{shift}$). As the spectrum was simulated with a simplified model and the existing hyperfine structure (line splitting $\sim0.6\,\mathrm{kHz}$ for our experimental parameters~\cite{Fabricant1977,Muller2017}) was neglected, we estimate as a conservative value that the Stark-shifted line position of the transition $\ket{4,3,2}\leftrightarrow\ket{5,3,3}$ is measured with an accuracy of at least a couple of hundred Hz. 

The transition frequency without electric field can be computed from the measurement considering the calculated minimum of the differential Stark shift, $(172.5\pm3.1)\,\mathrm{kHz}$, which depends on the rotational constants and is thus limited by their accuracy~\cite{Brunken2003}. Consequently, the zero-field center frequency is deduced with the same uncertainty as $(364\,282\,010\pm3)\,\mathrm{kHz}$ \footnote{With the proof-of-concept demonstration of our method, the uncertainty of the zero-field center frequency could be reduced by a factor of $\sim$\,10 compared to the most precise direct measurement reported in the literature~\cite{Cornet1980}.}.

The precision can be further increased and Doppler broadening reduced by using colder molecules. However, with the present experimental sequence and apparatus, the coldest and thus slowest ensembles of molecules we are currently able to produce in our setup ($T\sim400\,\mathrm{\mu K}$~\cite{Prehn2016}) cannot be probed in the comparably high \enquote*{magic} field of 1.05\,kV/cm required for the selected \enquote*{clock} transition $\ket{4,3,2}\leftrightarrow\ket{5,3,3}$. The reason is the finite roughness $\sigma_\mathrm{offset}$ of the floor of our boxlike trap potential [\reffig{fig:trap}(b)] which depends on both the high trapping fields and the offset field~\cite{Zeppenfeld2013}. To reduce the roughness of the floor while the molecules are further cooled, we start the cooling process at an offset electric field of 625\,V/cm and ramp the field down to as low as 40\,V/cm to produce molecules at submillikelvin temperatures~\cite{Prehn2016}. In fact, the kinetic energy of 400\,$\mu$K molecules does not exceed the roughness of the \enquote*{magic} offset potential. Hence, a well-defined \enquote*{magic} offset potential could not be applied for such ultracold molecules. When we adjust the offset field to the \enquote*{magic} value, very slow molecules are either heated or accumulate in the low-field regions ($\mathcal{E}<\mathcal{E}_\mathrm{offset}$) of the trap, precluding the investigation of the slowest molecules in a high offset field. We note that the median kinetic energy of the coldest ensemble used for spectroscopy ($E_\mathrm{kin}\sim\frac{3}{2}k_B\times2.6\,\mathrm{mK}$) is already a factor of 10 smaller than the potential energy of the molecules given by the floor of our boxlike trap potential ($\sim\frac{3}{2}k_B\times30\,\mathrm{mK}$ for the state $\ket{3,3,3}$ and $\mathcal{E}_\mathrm{offset}=1.05\,\mathrm{kV/cm}$). In consequence, the tunability of the relevant parameters, molecular velocity and offset electric field, is limited for purely technical reasons at the moment. These limitations could be overcome with an improved design of the trap electrodes, reducing the relative roughness of the offset electric fields and enabling cooling to submillikelvin temperatures at high offset fields~\cite{Zeppenfeld2013}. 

More fundamentally, the selection of a different \enquote*{clock} transition giving rise to a significantly reduced \enquote*{magic} field would allow probing colder molecules without redesigning the electric trap. Furthermore, it would permit choosing transitions with or without hyperfine splitting. For example, the \enquote*{magic} field is 36\,V/cm for $\ket{4,4,2}\leftrightarrow\ket{5,4,3}$ (no hyperfine splitting), while it is 140\,V/cm for $\ket{6,5,3}\leftrightarrow\ket{7,5,4}$ (with hyperfine splitting). However, these states are not accessible with our current implementation of optoelectrical Sisyphus cooling.

To summarize, Doppler-limited linewidths down to 3.8\,kHz were observed, yielding a new record for electrically trapped polyatomic molecules. The work demonstrates the advantage of our design over previous electric traps: Stark broadening could be reduced to a minimum due to the tunable, boxlike trapping potential. The presented method for high-resolution spectroscopy is general and applicable to other electrically trappable molecule species. Moreover, the results show the feasibility of precision spectroscopy in electric traps, a straightforward approach to trapping polar molecules.

In the future, it should be possible to resolve the hyperfine structure of formaldehyde with sub-kHz line splittings (in a moderately strong electric field)~\cite{Fabricant1977,Muller2017}. With colder molecules we should be able to observe Rabi oscillations in a two-level system coupled by a \enquote*{clock} transition, a key step towards coherent control over internal molecular states. The combination of box-like trapping fields with extremely long storage times of cooled polar molecules also offers new possibilities for the investigation of weak forbidden transitions.


\begin{thebibliography}{42}%
	\makeatletter
	\providecommand \@ifxundefined [1]{%
		\@ifx{#1\undefined}
	}%
	\providecommand \@ifnum [1]{%
		\ifnum #1\expandafter \@firstoftwo
		\else \expandafter \@secondoftwo
		\fi
	}%
	\providecommand \@ifx [1]{%
		\ifx #1\expandafter \@firstoftwo
		\else \expandafter \@secondoftwo
		\fi
	}%
	\providecommand \natexlab [1]{#1}%
	\providecommand \enquote  [1]{``#1''}%
	\providecommand \bibnamefont  [1]{#1}%
	\providecommand \bibfnamefont [1]{#1}%
	\providecommand \citenamefont [1]{#1}%
	\providecommand \href@noop [0]{\@secondoftwo}%
	\providecommand \href [0]{\begingroup \@sanitize@url \@href}%
	\providecommand \@href[1]{\@@startlink{#1}\@@href}%
	\providecommand \@@href[1]{\endgroup#1\@@endlink}%
	\providecommand \@sanitize@url [0]{\catcode `\\12\catcode `\$12\catcode
		`\&12\catcode `\#12\catcode `\^12\catcode `\_12\catcode `\%12\relax}%
	\providecommand \@@startlink[1]{}%
	\providecommand \@@endlink[0]{}%
	\providecommand \url  [0]{\begingroup\@sanitize@url \@url }%
	\providecommand \@url [1]{\endgroup\@href {#1}{\urlprefix }}%
	\providecommand \urlprefix  [0]{URL }%
	\providecommand \Eprint [0]{\href }%
	\providecommand \doibase [0]{http://dx.doi.org/}%
	\providecommand \selectlanguage [0]{\@gobble}%
	\providecommand \bibinfo  [0]{\@secondoftwo}%
	\providecommand \bibfield  [0]{\@secondoftwo}%
	\providecommand \translation [1]{[#1]}%
	\providecommand \BibitemOpen [0]{}%
	\providecommand \bibitemStop [0]{}%
	\providecommand \bibitemNoStop [0]{.\EOS\space}%
	\providecommand \EOS [0]{\spacefactor3000\relax}%
	\providecommand \BibitemShut  [1]{\csname bibitem#1\endcsname}%
	\let\auto@bib@innerbib\@empty
	\bibitem [{\citenamefont {Rosenband}\ \emph {et~al.}(2008)\citenamefont
		{Rosenband}, \citenamefont {Hume}, \citenamefont {Schmidt}, \citenamefont
		{Chou}, \citenamefont {Brusch}, \citenamefont {Lorini}, \citenamefont
		{Oskay}, \citenamefont {Drullinger}, \citenamefont {Fortier}, \citenamefont
		{Stalnaker}, \citenamefont {Diddams}, \citenamefont {Swann}, \citenamefont
		{Newbury}, \citenamefont {Itano}, \citenamefont {Wineland},\ and\
		\citenamefont {Bergquist}}]{Rosenband2008}%
	\BibitemOpen
	\bibfield  {author} {\bibinfo {author} {\bibfnamefont {T.}~\bibnamefont
			{Rosenband}}, \bibinfo {author} {\bibfnamefont {D.~B.}\ \bibnamefont {Hume}},
		\bibinfo {author} {\bibfnamefont {P.~O.}\ \bibnamefont {Schmidt}}, \bibinfo
		{author} {\bibfnamefont {C.~W.}\ \bibnamefont {Chou}}, \bibinfo {author}
		{\bibfnamefont {A.}~\bibnamefont {Brusch}}, \bibinfo {author} {\bibfnamefont
			{L.}~\bibnamefont {Lorini}}, \bibinfo {author} {\bibfnamefont {W.~H.}\
			\bibnamefont {Oskay}}, \bibinfo {author} {\bibfnamefont {R.~E.}\ \bibnamefont
			{Drullinger}}, \bibinfo {author} {\bibfnamefont {T.~M.}\ \bibnamefont
			{Fortier}}, \bibinfo {author} {\bibfnamefont {J.~E.}\ \bibnamefont
			{Stalnaker}}, \bibinfo {author} {\bibfnamefont {S.~A.}\ \bibnamefont
			{Diddams}}, \bibinfo {author} {\bibfnamefont {W.~C.}\ \bibnamefont {Swann}},
		\bibinfo {author} {\bibfnamefont {N.~R.}\ \bibnamefont {Newbury}}, \bibinfo
		{author} {\bibfnamefont {W.~M.}\ \bibnamefont {Itano}}, \bibinfo {author}
		{\bibfnamefont {D.~J.}\ \bibnamefont {Wineland}}, \ and\ \bibinfo {author}
		{\bibfnamefont {J.~C.}\ \bibnamefont {Bergquist}},\ }\bibfield  {title}
	{\enquote {\bibinfo {title} {{Frequency Ratio of Al$^+$ and Hg$^+$ Single-Ion
					Optical Clocks; Metrology at the 17th Decimal Place}},}\ }\href {\doibase
		10.1126/science.1154622} {\bibfield  {journal} {\bibinfo  {journal}
			{Science}\ }\textbf {\bibinfo {volume} {319}},\ \bibinfo {pages} {1808}
		(\bibinfo {year} {2008})}\BibitemShut {NoStop}%
	\bibitem [{\citenamefont {Ludlow}\ \emph {et~al.}(2015)\citenamefont {Ludlow},
		\citenamefont {Boyd}, \citenamefont {Ye}, \citenamefont {Peik},\ and\
		\citenamefont {Schmidt}}]{Ludlow2015}%
	\BibitemOpen
	\bibfield  {author} {\bibinfo {author} {\bibfnamefont {A.~D.}\ \bibnamefont
			{Ludlow}}, \bibinfo {author} {\bibfnamefont {M.~M.}\ \bibnamefont {Boyd}},
		\bibinfo {author} {\bibfnamefont {J.}~\bibnamefont {Ye}}, \bibinfo {author}
		{\bibfnamefont {E.}~\bibnamefont {Peik}}, \ and\ \bibinfo {author}
		{\bibfnamefont {P.~O.}\ \bibnamefont {Schmidt}},\ }\bibfield  {title}
	{\enquote {\bibinfo {title} {{Optical atomic clocks}},}\ }\href {\doibase
		10.1103/RevModPhys.87.637} {\bibfield  {journal} {\bibinfo  {journal} {Rev.
				Mod. Phys.}\ }\textbf {\bibinfo {volume} {87}},\ \bibinfo {pages} {637}
		(\bibinfo {year} {2015})}\BibitemShut {NoStop}%
	\bibitem [{\citenamefont {Griffith}\ \emph {et~al.}(2009)\citenamefont
		{Griffith}, \citenamefont {Swallows}, \citenamefont {Loftus}, \citenamefont
		{Romalis}, \citenamefont {Heckel},\ and\ \citenamefont
		{Fortson}}]{Griffith2009}%
	\BibitemOpen
	\bibfield  {author} {\bibinfo {author} {\bibfnamefont {W.~C.}\ \bibnamefont
			{Griffith}}, \bibinfo {author} {\bibfnamefont {M.~D.}\ \bibnamefont
			{Swallows}}, \bibinfo {author} {\bibfnamefont {T.~H.}\ \bibnamefont
			{Loftus}}, \bibinfo {author} {\bibfnamefont {M.~V.}\ \bibnamefont {Romalis}},
		\bibinfo {author} {\bibfnamefont {B.~R.}\ \bibnamefont {Heckel}}, \ and\
		\bibinfo {author} {\bibfnamefont {E.~N.}\ \bibnamefont {Fortson}},\
	}\bibfield  {title} {\enquote {\bibinfo {title} {{Improved Limit on the
					Permanent Electric Dipole Moment of $^{199}$Hg}},}\ }\href {\doibase
		10.1103/PhysRevLett.102.101601} {\bibfield  {journal} {\bibinfo  {journal}
			{Phys. Rev. Lett.}\ }\textbf {\bibinfo {volume} {102}},\ \bibinfo {pages}
		{101601} (\bibinfo {year} {2009})}\BibitemShut {NoStop}%
	\bibitem [{\citenamefont {Steimle}\ and\ \citenamefont
		{Ubachs}(2014)}]{Steimle2014}%
	\BibitemOpen
	\bibfield  {author} {\bibinfo {author} {\bibfnamefont {T.}~\bibnamefont
			{Steimle}}\ and\ \bibinfo {author} {\bibfnamefont {W.}~\bibnamefont
			{Ubachs}},\ }\bibfield  {title} {\enquote {\bibinfo {title} {{Introduction to
					the special issue spectroscopic tests of fundamental physics}},}\ }\href
	{\doibase 10.1016/j.jms.2014.04.004} {\bibfield  {journal} {\bibinfo
			{journal} {J. Mol. Spectrosc.}\ }\textbf {\bibinfo {volume} {300}},\ \bibinfo
		{pages} {1} (\bibinfo {year} {2014})}\BibitemShut {NoStop}%
	\bibitem [{\citenamefont {DeMille}\ \emph {et~al.}(2017)\citenamefont
		{DeMille}, \citenamefont {Doyle},\ and\ \citenamefont
		{Sushkov}}]{DeMille2017}%
	\BibitemOpen
	\bibfield  {author} {\bibinfo {author} {\bibfnamefont {D.}~\bibnamefont
			{DeMille}}, \bibinfo {author} {\bibfnamefont {J.~M.}\ \bibnamefont {Doyle}},
		\ and\ \bibinfo {author} {\bibfnamefont {A.~O.}\ \bibnamefont {Sushkov}},\
	}\bibfield  {title} {\enquote {\bibinfo {title} {{Probing the frontiers of
					particle physics with tabletop-scale experiments}},}\ }\href {\doibase
		10.1126/science.aal3003} {\bibfield  {journal} {\bibinfo  {journal}
			{Science}\ }\textbf {\bibinfo {volume} {357}},\ \bibinfo {pages} {990}
		(\bibinfo {year} {2017})}\BibitemShut {NoStop}%
	\bibitem [{\citenamefont {Baron}\ \emph {et~al.}(2014)\citenamefont {Baron},
		\citenamefont {Campbell}, \citenamefont {DeMille}, \citenamefont {Doyle},
		\citenamefont {Gabrielse}, \citenamefont {Gurevich}, \citenamefont {Hess},
		\citenamefont {Hutzler}, \citenamefont {Kirilov}, \citenamefont {Kozyryev},
		\citenamefont {O'Leary}, \citenamefont {Panda}, \citenamefont {Parsons},
		\citenamefont {Petrik}, \citenamefont {Spaun}, \citenamefont {Vutha},\ and\
		\citenamefont {West}}]{Baron2014}%
	\BibitemOpen
	\bibfield  {author} {\bibinfo {author} {\bibfnamefont {J.}~\bibnamefont
			{Baron}}, \bibinfo {author} {\bibfnamefont {W.~C.}\ \bibnamefont {Campbell}},
		\bibinfo {author} {\bibfnamefont {D.}~\bibnamefont {DeMille}}, \bibinfo
		{author} {\bibfnamefont {J.~M.}\ \bibnamefont {Doyle}}, \bibinfo {author}
		{\bibfnamefont {G.}~\bibnamefont {Gabrielse}}, \bibinfo {author}
		{\bibfnamefont {Y.~V.}\ \bibnamefont {Gurevich}}, \bibinfo {author}
		{\bibfnamefont {P.~W.}\ \bibnamefont {Hess}}, \bibinfo {author}
		{\bibfnamefont {N.~R.}\ \bibnamefont {Hutzler}}, \bibinfo {author}
		{\bibfnamefont {E.}~\bibnamefont {Kirilov}}, \bibinfo {author} {\bibfnamefont
			{I.}~\bibnamefont {Kozyryev}}, \bibinfo {author} {\bibfnamefont {B.~R.}\
			\bibnamefont {O'Leary}}, \bibinfo {author} {\bibfnamefont {C.~D.}\
			\bibnamefont {Panda}}, \bibinfo {author} {\bibfnamefont {M.~F.}\ \bibnamefont
			{Parsons}}, \bibinfo {author} {\bibfnamefont {E.~S.}\ \bibnamefont {Petrik}},
		\bibinfo {author} {\bibfnamefont {B.}~\bibnamefont {Spaun}}, \bibinfo
		{author} {\bibfnamefont {A.~C.}\ \bibnamefont {Vutha}}, \ and\ \bibinfo
		{author} {\bibfnamefont {A.~D.}\ \bibnamefont {West}},\ }\bibfield  {title}
	{\enquote {\bibinfo {title} {{Order of Magnitude Smaller Limit on the
					Electric Dipole Moment of the Electron}},}\ }\href {\doibase
		10.1126/science.1248213} {\bibfield  {journal} {\bibinfo  {journal}
			{Science}\ }\textbf {\bibinfo {volume} {343}},\ \bibinfo {pages} {269}
		(\bibinfo {year} {2014})}\BibitemShut {NoStop}%
	\bibitem [{\citenamefont {Cairncross}\ \emph {et~al.}(2017)\citenamefont
		{Cairncross}, \citenamefont {Gresh}, \citenamefont {Grau}, \citenamefont
		{Cossel}, \citenamefont {Roussy}, \citenamefont {Ni}, \citenamefont {Zhou},
		\citenamefont {Ye},\ and\ \citenamefont {Cornell}}]{Cairncross2017}%
	\BibitemOpen
	\bibfield  {author} {\bibinfo {author} {\bibfnamefont {W.~B.}\ \bibnamefont
			{Cairncross}}, \bibinfo {author} {\bibfnamefont {D.~N.}\ \bibnamefont
			{Gresh}}, \bibinfo {author} {\bibfnamefont {M.}~\bibnamefont {Grau}},
		\bibinfo {author} {\bibfnamefont {K.~C.}\ \bibnamefont {Cossel}}, \bibinfo
		{author} {\bibfnamefont {T.~S.}\ \bibnamefont {Roussy}}, \bibinfo {author}
		{\bibfnamefont {Y.}~\bibnamefont {Ni}}, \bibinfo {author} {\bibfnamefont
			{Y.}~\bibnamefont {Zhou}}, \bibinfo {author} {\bibfnamefont {J.}~\bibnamefont
			{Ye}}, \ and\ \bibinfo {author} {\bibfnamefont {E.~A.}\ \bibnamefont
			{Cornell}},\ }\bibfield  {title} {\enquote {\bibinfo {title} {{Precision
					Measurement of the Electron's Electric Dipole Moment Using Trapped Molecular
					Ions}},}\ }\href {\doibase 10.1103/PhysRevLett.119.153001} {\bibfield
		{journal} {\bibinfo  {journal} {Phys. Rev. Lett.}\ }\textbf {\bibinfo
			{volume} {119}},\ \bibinfo {pages} {153001} (\bibinfo {year}
		{2017})}\BibitemShut {NoStop}%
	\bibitem [{\citenamefont {Quack}(2011)}]{Quack2011}%
	\BibitemOpen
	\bibfield  {author} {\bibinfo {author} {\bibfnamefont {M.}~\bibnamefont
			{Quack}},\ }\bibfield  {title} {\enquote {\bibinfo {title} {{Frontiers in
					spectroscopy}},}\ }\href {\doibase 10.1039/c1fd00096a} {\bibfield  {journal}
		{\bibinfo  {journal} {Faraday Discuss.}\ }\textbf {\bibinfo {volume} {150}},\
		\bibinfo {pages} {533} (\bibinfo {year} {2011})}\BibitemShut {NoStop}%
	\bibitem [{\citenamefont {Mejri}\ \emph {et~al.}(2015)\citenamefont {Mejri},
		\citenamefont {Sow}, \citenamefont {Kozlova}, \citenamefont {Ayari},
		\citenamefont {Tokunaga}, \citenamefont {Chardonnet}, \citenamefont
		{Briaudeau}, \citenamefont {Darqui{\'{e}}}, \citenamefont {Rohart},\ and\
		\citenamefont {Daussy}}]{Mejri2015}%
	\BibitemOpen
	\bibfield  {author} {\bibinfo {author} {\bibfnamefont {S.}~\bibnamefont
			{Mejri}}, \bibinfo {author} {\bibfnamefont {P.~L.~T.}\ \bibnamefont {Sow}},
		\bibinfo {author} {\bibfnamefont {O.}~\bibnamefont {Kozlova}}, \bibinfo
		{author} {\bibfnamefont {C.}~\bibnamefont {Ayari}}, \bibinfo {author}
		{\bibfnamefont {S.~K.}\ \bibnamefont {Tokunaga}}, \bibinfo {author}
		{\bibfnamefont {C.}~\bibnamefont {Chardonnet}}, \bibinfo {author}
		{\bibfnamefont {S.}~\bibnamefont {Briaudeau}}, \bibinfo {author}
		{\bibfnamefont {B.}~\bibnamefont {Darqui{\'{e}}}}, \bibinfo {author}
		{\bibfnamefont {F.}~\bibnamefont {Rohart}}, \ and\ \bibinfo {author}
		{\bibfnamefont {C.}~\bibnamefont {Daussy}},\ }\bibfield  {title} {\enquote
		{\bibinfo {title} {{Measuring the Boltzmann constant by mid-infrared laser
					spectroscopy of ammonia}},}\ }\href {\doibase 10.1088/0026-1394/52/5/S314}
	{\bibfield  {journal} {\bibinfo  {journal} {Metrologia}\ }\textbf {\bibinfo
			{volume} {52}},\ \bibinfo {pages} {S314} (\bibinfo {year}
		{2015})}\BibitemShut {NoStop}%
	\bibitem [{\citenamefont {Biesheuvel}\ \emph {et~al.}(2016)\citenamefont
		{Biesheuvel}, \citenamefont {Karr}, \citenamefont {Hilico}, \citenamefont
		{Eikema}, \citenamefont {Ubachs},\ and\ \citenamefont
		{Koelemeij}}]{Biesheuvel2016}%
	\BibitemOpen
	\bibfield  {author} {\bibinfo {author} {\bibfnamefont {J.}~\bibnamefont
			{Biesheuvel}}, \bibinfo {author} {\bibfnamefont {J.~P.}\ \bibnamefont
			{Karr}}, \bibinfo {author} {\bibfnamefont {L.}~\bibnamefont {Hilico}},
		\bibinfo {author} {\bibfnamefont {K.~S.~E.}\ \bibnamefont {Eikema}}, \bibinfo
		{author} {\bibfnamefont {W.}~\bibnamefont {Ubachs}}, \ and\ \bibinfo {author}
		{\bibfnamefont {J.~C.~J.}\ \bibnamefont {Koelemeij}},\ }\bibfield  {title}
	{\enquote {\bibinfo {title} {{Probing QED and fundamental constants through
					laser spectroscopy of vibrational transitions in HD$^+$}},}\ }\href {\doibase
		10.1038/ncomms10385} {\bibfield  {journal} {\bibinfo  {journal} {Nat.
				Commun.}\ }\textbf {\bibinfo {volume} {7}},\ \bibinfo {pages} {10385}
		(\bibinfo {year} {2016})}\BibitemShut {NoStop}%
	\bibitem [{\citenamefont {Shelkovnikov}\ \emph {et~al.}(2008)\citenamefont
		{Shelkovnikov}, \citenamefont {Butcher}, \citenamefont {Chardonnet},\ and\
		\citenamefont {Amy-Klein}}]{Shelkovnikov2008}%
	\BibitemOpen
	\bibfield  {author} {\bibinfo {author} {\bibfnamefont {A.}~\bibnamefont
			{Shelkovnikov}}, \bibinfo {author} {\bibfnamefont {R.~J.}\ \bibnamefont
			{Butcher}}, \bibinfo {author} {\bibfnamefont {C.}~\bibnamefont {Chardonnet}},
		\ and\ \bibinfo {author} {\bibfnamefont {A.}~\bibnamefont {Amy-Klein}},\
	}\bibfield  {title} {\enquote {\bibinfo {title} {{Stability of the
					Proton-to-Electron Mass Ratio}},}\ }\href {\doibase
		10.1103/PhysRevLett.100.150801} {\bibfield  {journal} {\bibinfo  {journal}
			{Phys. Rev. Lett.}\ }\textbf {\bibinfo {volume} {100}},\ \bibinfo {pages}
		{150801} (\bibinfo {year} {2008})}\BibitemShut {NoStop}%
	\bibitem [{\citenamefont {Truppe}\ \emph {et~al.}(2013)\citenamefont {Truppe},
		\citenamefont {Hendricks}, \citenamefont {Tokunaga}, \citenamefont
		{Lewandowski}, \citenamefont {Kozlov}, \citenamefont {Henkel}, \citenamefont
		{Hinds},\ and\ \citenamefont {Tarbutt}}]{Truppe2013}%
	\BibitemOpen
	\bibfield  {author} {\bibinfo {author} {\bibfnamefont {S.}~\bibnamefont
			{Truppe}}, \bibinfo {author} {\bibfnamefont {R.~J.}\ \bibnamefont
			{Hendricks}}, \bibinfo {author} {\bibfnamefont {S.~K.}\ \bibnamefont
			{Tokunaga}}, \bibinfo {author} {\bibfnamefont {H.~J.}\ \bibnamefont
			{Lewandowski}}, \bibinfo {author} {\bibfnamefont {M.~G.}\ \bibnamefont
			{Kozlov}}, \bibinfo {author} {\bibfnamefont {C.}~\bibnamefont {Henkel}},
		\bibinfo {author} {\bibfnamefont {E.~A.}\ \bibnamefont {Hinds}}, \ and\
		\bibinfo {author} {\bibfnamefont {M.~R.}\ \bibnamefont {Tarbutt}},\
	}\bibfield  {title} {\enquote {\bibinfo {title} {{A search for varying
					fundamental constants using hertz-level frequency measurements of cold CH
					molecules.}}}\ }\href {\doibase 10.1038/ncomms3600} {\bibfield  {journal}
		{\bibinfo  {journal} {Nat. Commun.}\ }\textbf {\bibinfo {volume} {4}},\
		\bibinfo {pages} {2600} (\bibinfo {year} {2013})}\BibitemShut {NoStop}%
	\bibitem [{\citenamefont {Bell}\ and\ \citenamefont {{P.
				Softley}}(2009)}]{Bell2009}%
	\BibitemOpen
	\bibfield  {author} {\bibinfo {author} {\bibfnamefont {M.~T.}\ \bibnamefont
			{Bell}}\ and\ \bibinfo {author} {\bibfnamefont {T.}~\bibnamefont {{P.
					Softley}}},\ }\bibfield  {title} {\enquote {\bibinfo {title} {{Ultracold
					molecules and ultracold chemistry}},}\ }\href {\doibase
		10.1080/00268970902724955} {\bibfield  {journal} {\bibinfo  {journal} {Mol.
				Phys.}\ }\textbf {\bibinfo {volume} {107}},\ \bibinfo {pages} {99} (\bibinfo
		{year} {2009})}\BibitemShut {NoStop}%
	\bibitem [{\citenamefont {Bohn}\ \emph {et~al.}(2017)\citenamefont {Bohn},
		\citenamefont {Rey},\ and\ \citenamefont {Ye}}]{Bohn2017}%
	\BibitemOpen
	\bibfield  {author} {\bibinfo {author} {\bibfnamefont {J.~L.}\ \bibnamefont
			{Bohn}}, \bibinfo {author} {\bibfnamefont {A.~M.}\ \bibnamefont {Rey}}, \
		and\ \bibinfo {author} {\bibfnamefont {J.}~\bibnamefont {Ye}},\ }\bibfield
	{title} {\enquote {\bibinfo {title} {{Cold molecules: Progress in quantum
					engineering of chemistry and quantum matter}},}\ }\href {\doibase
		10.1126/science.aam6299} {\bibfield  {journal} {\bibinfo  {journal}
			{Science}\ }\textbf {\bibinfo {volume} {357}},\ \bibinfo {pages} {1002}
		(\bibinfo {year} {2017})}\BibitemShut {NoStop}%
	\bibitem [{\citenamefont {Bergin}\ and\ \citenamefont
		{Tafalla}(2007)}]{Bergin2007}%
	\BibitemOpen
	\bibfield  {author} {\bibinfo {author} {\bibfnamefont {E.~A.}\ \bibnamefont
			{Bergin}}\ and\ \bibinfo {author} {\bibfnamefont {M.}~\bibnamefont
			{Tafalla}},\ }\bibfield  {title} {\enquote {\bibinfo {title} {{Cold Dark
					Clouds: The Initial Conditions for Star Formation}},}\ }\href {\doibase
		10.1146/annurev.astro.45.071206.100404} {\bibfield  {journal} {\bibinfo
			{journal} {Annu. Rev. Astron. Astrophys.}\ }\textbf {\bibinfo {volume}
			{45}},\ \bibinfo {pages} {339} (\bibinfo {year} {2007})}\BibitemShut
	{NoStop}%
	\bibitem [{\citenamefont {M{\"{u}}ller}\ and\ \citenamefont
		{Lewen}(2017)}]{Muller2017}%
	\BibitemOpen
	\bibfield  {author} {\bibinfo {author} {\bibfnamefont {H.~S.}\ \bibnamefont
			{M{\"{u}}ller}}\ and\ \bibinfo {author} {\bibfnamefont {F.}~\bibnamefont
			{Lewen}},\ }\bibfield  {title} {\enquote {\bibinfo {title} {{Submillimeter
					spectroscopy of H$_2$C$^{17}$O and a revisit of the rotational spectra of
					H$_2$C$^{18}$O and H$_2$C$^{16}$O}},}\ }\href {\doibase
		10.1016/j.jms.2016.10.004} {\bibfield  {journal} {\bibinfo  {journal} {J.
				Mol. Spectrosc.}\ }\textbf {\bibinfo {volume} {331}},\ \bibinfo {pages} {28}
		(\bibinfo {year} {2017})}\BibitemShut {NoStop}%
	\bibitem [{\citenamefont {Veldhoven}\ \emph {et~al.}(2004)\citenamefont
		{Veldhoven}, \citenamefont {K{\"{u}}pper}, \citenamefont {Bethlem},
		\citenamefont {Sartakov}, \citenamefont {Roij},\ and\ \citenamefont
		{Meijer}}]{Veldhoven2004}%
	\BibitemOpen
	\bibfield  {author} {\bibinfo {author} {\bibfnamefont {J.}~\bibnamefont
			{Veldhoven}}, \bibinfo {author} {\bibfnamefont {J.}~\bibnamefont
			{K{\"{u}}pper}}, \bibinfo {author} {\bibfnamefont {H.~L.}\ \bibnamefont
			{Bethlem}}, \bibinfo {author} {\bibfnamefont {B.}~\bibnamefont {Sartakov}},
		\bibinfo {author} {\bibfnamefont {A.~J.~A.}\ \bibnamefont {Roij}}, \ and\
		\bibinfo {author} {\bibfnamefont {G.}~\bibnamefont {Meijer}},\ }\bibfield
	{title} {\enquote {\bibinfo {title} {{Decelerated molecular beams for
					high-resolution spectroscopy}},}\ }\href {\doibase
		10.1140/epjd/e2004-00160-9} {\bibfield  {journal} {\bibinfo  {journal} {Eur.
				Phys. J. D}\ }\textbf {\bibinfo {volume} {31}},\ \bibinfo {pages} {337}
		(\bibinfo {year} {2004})}\BibitemShut {NoStop}%
	\bibitem [{\citenamefont {Hudson}\ \emph {et~al.}(2006)\citenamefont {Hudson},
		\citenamefont {Lewandowski}, \citenamefont {Sawyer},\ and\ \citenamefont
		{Ye}}]{Hudson2006a}%
	\BibitemOpen
	\bibfield  {author} {\bibinfo {author} {\bibfnamefont {E.~R.}\ \bibnamefont
			{Hudson}}, \bibinfo {author} {\bibfnamefont {H.~J.}\ \bibnamefont
			{Lewandowski}}, \bibinfo {author} {\bibfnamefont {B.~C.}\ \bibnamefont
			{Sawyer}}, \ and\ \bibinfo {author} {\bibfnamefont {J.}~\bibnamefont {Ye}},\
	}\bibfield  {title} {\enquote {\bibinfo {title} {{Cold Molecule Spectroscopy
					for Constraining the Evolution of the Fine Structure Constant}},}\ }\href
	{\doibase 10.1103/PhysRevLett.96.143004} {\bibfield  {journal} {\bibinfo
			{journal} {Phys. Rev. Lett.}\ }\textbf {\bibinfo {volume} {96}},\ \bibinfo
		{pages} {143004} (\bibinfo {year} {2006})}\BibitemShut {NoStop}%
	\bibitem [{\citenamefont {de~Nijs}\ \emph {et~al.}(2014)\citenamefont
		{de~Nijs}, \citenamefont {Ubachs},\ and\ \citenamefont
		{Bethlem}}]{DeNijs2014}%
	\BibitemOpen
	\bibfield  {author} {\bibinfo {author} {\bibfnamefont {A.}~\bibnamefont
			{de~Nijs}}, \bibinfo {author} {\bibfnamefont {W.}~\bibnamefont {Ubachs}}, \
		and\ \bibinfo {author} {\bibfnamefont {H.}~\bibnamefont {Bethlem}},\
	}\bibfield  {title} {\enquote {\bibinfo {title} {{Ramsey-type microwave
					spectroscopy on CO (a$^{3}\Pi$)}},}\ }\href {\doibase
		10.1016/j.jms.2014.03.020} {\bibfield  {journal} {\bibinfo  {journal} {J.
				Mol. Spectrosc.}\ }\textbf {\bibinfo {volume} {300}},\ \bibinfo {pages} {79}
		(\bibinfo {year} {2014})}\BibitemShut {NoStop}%
	\bibitem [{\citenamefont {Cahn}\ \emph {et~al.}(2014)\citenamefont {Cahn},
		\citenamefont {Ammon}, \citenamefont {Kirilov}, \citenamefont {Gurevich},
		\citenamefont {Murphree}, \citenamefont {Paolino}, \citenamefont {Rahmlow},
		\citenamefont {Kozlov},\ and\ \citenamefont {DeMille}}]{Cahn2014}%
	\BibitemOpen
	\bibfield  {author} {\bibinfo {author} {\bibfnamefont {S.~B.}\ \bibnamefont
			{Cahn}}, \bibinfo {author} {\bibfnamefont {J.}~\bibnamefont {Ammon}},
		\bibinfo {author} {\bibfnamefont {E.}~\bibnamefont {Kirilov}}, \bibinfo
		{author} {\bibfnamefont {Y.~V.}\ \bibnamefont {Gurevich}}, \bibinfo {author}
		{\bibfnamefont {D.}~\bibnamefont {Murphree}}, \bibinfo {author}
		{\bibfnamefont {R.}~\bibnamefont {Paolino}}, \bibinfo {author} {\bibfnamefont
			{D.~A.}\ \bibnamefont {Rahmlow}}, \bibinfo {author} {\bibfnamefont {M.~G.}\
			\bibnamefont {Kozlov}}, \ and\ \bibinfo {author} {\bibfnamefont
			{D.}~\bibnamefont {DeMille}},\ }\bibfield  {title} {\enquote {\bibinfo
			{title} {Zeeman-tuned rotational level-crossing spectroscopy in a diatomic
				free radical},}\ }\href {\doibase 10.1103/PhysRevLett.112.163002} {\bibfield
		{journal} {\bibinfo  {journal} {Phys. Rev. Lett.}\ }\textbf {\bibinfo
			{volume} {112}},\ \bibinfo {pages} {163002} (\bibinfo {year}
		{2014})}\BibitemShut {NoStop}%
	\bibitem [{\citenamefont {Bethlem}\ \emph {et~al.}(2008)\citenamefont
		{Bethlem}, \citenamefont {Kajita}, \citenamefont {Sartakov}, \citenamefont
		{Meijer},\ and\ \citenamefont {Ubachs}}]{Bethlem2008}%
	\BibitemOpen
	\bibfield  {author} {\bibinfo {author} {\bibfnamefont {H.~L.}\ \bibnamefont
			{Bethlem}}, \bibinfo {author} {\bibfnamefont {M.}~\bibnamefont {Kajita}},
		\bibinfo {author} {\bibfnamefont {B.}~\bibnamefont {Sartakov}}, \bibinfo
		{author} {\bibfnamefont {G.}~\bibnamefont {Meijer}}, \ and\ \bibinfo {author}
		{\bibfnamefont {W.}~\bibnamefont {Ubachs}},\ }\bibfield  {title} {\enquote
		{\bibinfo {title} {{Prospects for precision measurements on ammonia molecules
					in a fountain}},}\ }\href {\doibase 10.1140/epjst/e2008-00809-5} {\bibfield
		{journal} {\bibinfo  {journal} {Eur. Phys. J. Spec. Top.}\ }\textbf {\bibinfo
			{volume} {163}},\ \bibinfo {pages} {55} (\bibinfo {year} {2008})}\BibitemShut
	{NoStop}%
	\bibitem [{\citenamefont {Tarbutt}\ \emph {et~al.}(2013)\citenamefont
		{Tarbutt}, \citenamefont {Sauer}, \citenamefont {Hudson},\ and\ \citenamefont
		{Hinds}}]{Tarbutt2013}%
	\BibitemOpen
	\bibfield  {author} {\bibinfo {author} {\bibfnamefont {M.~R.}\ \bibnamefont
			{Tarbutt}}, \bibinfo {author} {\bibfnamefont {B.~E.}\ \bibnamefont {Sauer}},
		\bibinfo {author} {\bibfnamefont {J.~J.}\ \bibnamefont {Hudson}}, \ and\
		\bibinfo {author} {\bibfnamefont {E.~A.}\ \bibnamefont {Hinds}},\ }\bibfield
	{title} {\enquote {\bibinfo {title} {{Design for a fountain of YbF molecules
					to measure the electron's electric dipole moment}},}\ }\href {\doibase
		10.1088/1367-2630/15/5/053034} {\bibfield  {journal} {\bibinfo  {journal}
			{New J. Phys.}\ }\textbf {\bibinfo {volume} {15}},\ \bibinfo {pages} {053034}
		(\bibinfo {year} {2013})}\BibitemShut {NoStop}%
	\bibitem [{\citenamefont {Cheng}\ \emph {et~al.}(2016)\citenamefont {Cheng},
		\citenamefont {van~der Poel}, \citenamefont {Jansen}, \citenamefont
		{Quintero-P{\'{e}}rez}, \citenamefont {Wall}, \citenamefont {Ubachs},\ and\
		\citenamefont {Bethlem}}]{Cheng2016}%
	\BibitemOpen
	\bibfield  {author} {\bibinfo {author} {\bibfnamefont {C.}~\bibnamefont
			{Cheng}}, \bibinfo {author} {\bibfnamefont {A.~P.~P.}\ \bibnamefont {van~der
				Poel}}, \bibinfo {author} {\bibfnamefont {P.}~\bibnamefont {Jansen}},
		\bibinfo {author} {\bibfnamefont {M.}~\bibnamefont {Quintero-P{\'{e}}rez}},
		\bibinfo {author} {\bibfnamefont {T.~E.}\ \bibnamefont {Wall}}, \bibinfo
		{author} {\bibfnamefont {W.}~\bibnamefont {Ubachs}}, \ and\ \bibinfo {author}
		{\bibfnamefont {H.~L.}\ \bibnamefont {Bethlem}},\ }\bibfield  {title}
	{\enquote {\bibinfo {title} {{Molecular Fountain}},}\ }\href {\doibase
		10.1103/PhysRevLett.117.253201} {\bibfield  {journal} {\bibinfo  {journal}
			{Phys. Rev. Lett.}\ }\textbf {\bibinfo {volume} {117}},\ \bibinfo {pages}
		{253201} (\bibinfo {year} {2016})}\BibitemShut {NoStop}%
	\bibitem [{\citenamefont {Wall}(2016)}]{Wall2016}%
	\BibitemOpen
	\bibfield  {author} {\bibinfo {author} {\bibfnamefont {T.~E.}\ \bibnamefont
			{Wall}},\ }\bibfield  {title} {\enquote {\bibinfo {title} {{Preparation of
					cold molecules for high-precision measurements}},}\ }\href {\doibase
		10.1088/0953-4075/49/24/243001} {\bibfield  {journal} {\bibinfo  {journal}
			{J. Phys. B}\ }\textbf {\bibinfo {volume} {49}},\ \bibinfo {pages} {243001}
		(\bibinfo {year} {2016})}\BibitemShut {NoStop}%
	\bibitem [{\citenamefont {Kozyryev}\ and\ \citenamefont
		{Hutzler}(2017)}]{Kozyryev2017}%
	\BibitemOpen
	\bibfield  {author} {\bibinfo {author} {\bibfnamefont {I.}~\bibnamefont
			{Kozyryev}}\ and\ \bibinfo {author} {\bibfnamefont {N.~R.}\ \bibnamefont
			{Hutzler}},\ }\bibfield  {title} {\enquote {\bibinfo {title} {{Precision
					Measurement of Time-Reversal Symmetry Violation with Laser-Cooled Polyatomic
					Molecules}},}\ }\href {\doibase 10.1103/PhysRevLett.119.133002} {\bibfield
		{journal} {\bibinfo  {journal} {Phys. Rev. Lett.}\ }\textbf {\bibinfo
			{volume} {119}},\ \bibinfo {pages} {133002} (\bibinfo {year}
		{2017})}\BibitemShut {NoStop}%
	\bibitem [{\citenamefont {Alighanbari}\ \emph {et~al.}(2018)\citenamefont
		{Alighanbari}, \citenamefont {Hansen}, \citenamefont {Korobov},\ and\
		\citenamefont {Schiller}}]{Alighanbari2018}%
	\BibitemOpen
	\bibfield  {author} {\bibinfo {author} {\bibfnamefont {S.}~\bibnamefont
			{Alighanbari}}, \bibinfo {author} {\bibfnamefont {M.~G.}\ \bibnamefont
			{Hansen}}, \bibinfo {author} {\bibfnamefont {V.~I.}\ \bibnamefont {Korobov}},
		\ and\ \bibinfo {author} {\bibfnamefont {S.}~\bibnamefont {Schiller}},\
	}\bibfield  {title} {\enquote {\bibinfo {title} {{Rotational spectroscopy of
					cold and trapped molecular ions in the Lamb–Dicke regime}},}\ }\href
	{\doibase 10.1038/s41567-018-0074-3} {\bibfield  {journal} {\bibinfo
			{journal} {Nat. Phys.}\ }\textbf {\bibinfo {volume} {14}},\ \bibinfo {pages}
		{555} (\bibinfo {year} {2018})}\BibitemShut {NoStop}%
	\bibitem [{\citenamefont {McGuyer}\ \emph {et~al.}(2015)\citenamefont
		{McGuyer}, \citenamefont {McDonald}, \citenamefont {Iwata}, \citenamefont
		{Tarallo}, \citenamefont {Grier}, \citenamefont {Apfelbeck},\ and\
		\citenamefont {Zelevinsky}}]{McGuyer2015}%
	\BibitemOpen
	\bibfield  {author} {\bibinfo {author} {\bibfnamefont {B.~H.}\ \bibnamefont
			{McGuyer}}, \bibinfo {author} {\bibfnamefont {M.}~\bibnamefont {McDonald}},
		\bibinfo {author} {\bibfnamefont {G.~Z.}\ \bibnamefont {Iwata}}, \bibinfo
		{author} {\bibfnamefont {M.~G.}\ \bibnamefont {Tarallo}}, \bibinfo {author}
		{\bibfnamefont {A.~T.}\ \bibnamefont {Grier}}, \bibinfo {author}
		{\bibfnamefont {F.}~\bibnamefont {Apfelbeck}}, \ and\ \bibinfo {author}
		{\bibfnamefont {T.}~\bibnamefont {Zelevinsky}},\ }\bibfield  {title}
	{\enquote {\bibinfo {title} {{High-precision spectroscopy of ultracold
					molecules in an optical lattice}},}\ }\href {\doibase
		10.1088/1367-2630/17/5/055004} {\bibfield  {journal} {\bibinfo  {journal}
			{New J. Phys.}\ }\textbf {\bibinfo {volume} {17}},\ \bibinfo {pages} {055004}
		(\bibinfo {year} {2015})}\BibitemShut {NoStop}%
	\bibitem [{\citenamefont {Park}\ \emph {et~al.}(2017)\citenamefont {Park},
		\citenamefont {Yan}, \citenamefont {Loh}, \citenamefont {Will},\ and\
		\citenamefont {Zwierlein}}]{Park2017}%
	\BibitemOpen
	\bibfield  {author} {\bibinfo {author} {\bibfnamefont {J.~W.}\ \bibnamefont
			{Park}}, \bibinfo {author} {\bibfnamefont {Z.~Z.}\ \bibnamefont {Yan}},
		\bibinfo {author} {\bibfnamefont {H.}~\bibnamefont {Loh}}, \bibinfo {author}
		{\bibfnamefont {S.~A.}\ \bibnamefont {Will}}, \ and\ \bibinfo {author}
		{\bibfnamefont {M.~W.}\ \bibnamefont {Zwierlein}},\ }\bibfield  {title}
	{\enquote {\bibinfo {title} {{Second-scale nuclear spin coherence time of
					ultracold $^{23}$Na$^{40}$K molecules}},}\ }\href {\doibase
		10.1126/science.aal5066} {\bibfield  {journal} {\bibinfo  {journal}
			{Science}\ }\textbf {\bibinfo {volume} {357}},\ \bibinfo {pages} {372}
		(\bibinfo {year} {2017})}\BibitemShut {NoStop}%
	\bibitem [{\citenamefont {Prehn}\ \emph {et~al.}(2016)\citenamefont {Prehn},
		\citenamefont {Ibr{\"{u}}gger}, \citenamefont {Gl{\"{o}}ckner}, \citenamefont
		{Rempe},\ and\ \citenamefont {Zeppenfeld}}]{Prehn2016}%
	\BibitemOpen
	\bibfield  {author} {\bibinfo {author} {\bibfnamefont {A.}~\bibnamefont
			{Prehn}}, \bibinfo {author} {\bibfnamefont {M.}~\bibnamefont
			{Ibr{\"{u}}gger}}, \bibinfo {author} {\bibfnamefont {R.}~\bibnamefont
			{Gl{\"{o}}ckner}}, \bibinfo {author} {\bibfnamefont {G.}~\bibnamefont
			{Rempe}}, \ and\ \bibinfo {author} {\bibfnamefont {M.}~\bibnamefont
			{Zeppenfeld}},\ }\bibfield  {title} {\enquote {\bibinfo {title}
			{{Optoelectrical Cooling of Polar Molecules to Submillikelvin
					Temperatures}},}\ }\href {\doibase 10.1103/PhysRevLett.116.063005} {\bibfield
		{journal} {\bibinfo  {journal} {Phys. Rev. Lett.}\ }\textbf {\bibinfo
			{volume} {116}},\ \bibinfo {pages} {063005} (\bibinfo {year}
		{2016})}\BibitemShut {NoStop}%
	\bibitem [{\citenamefont {Norrgard}\ \emph {et~al.}(2016)\citenamefont
		{Norrgard}, \citenamefont {McCarron}, \citenamefont {Steinecker},
		\citenamefont {Tarbutt},\ and\ \citenamefont {DeMille}}]{Norrgard2016}%
	\BibitemOpen
	\bibfield  {author} {\bibinfo {author} {\bibfnamefont {E.~B.}\ \bibnamefont
			{Norrgard}}, \bibinfo {author} {\bibfnamefont {D.~J.}\ \bibnamefont
			{McCarron}}, \bibinfo {author} {\bibfnamefont {M.~H.}\ \bibnamefont
			{Steinecker}}, \bibinfo {author} {\bibfnamefont {M.~R.}\ \bibnamefont
			{Tarbutt}}, \ and\ \bibinfo {author} {\bibfnamefont {D.}~\bibnamefont
			{DeMille}},\ }\bibfield  {title} {\enquote {\bibinfo {title} {{Submillikelvin
					Dipolar Molecules in a Radio-Frequency Magneto-Optical Trap}},}\ }\href
	{\doibase 10.1103/PhysRevLett.116.063004} {\bibfield  {journal} {\bibinfo
			{journal} {Phys. Rev. Lett.}\ }\textbf {\bibinfo {volume} {116}},\ \bibinfo
		{pages} {063004} (\bibinfo {year} {2016})}\BibitemShut {NoStop}%
	\bibitem [{\citenamefont {Ye}\ \emph {et~al.}(2008)\citenamefont {Ye},
		\citenamefont {Kimble},\ and\ \citenamefont {Katori}}]{Ye2008}%
	\BibitemOpen
	\bibfield  {author} {\bibinfo {author} {\bibfnamefont {J.}~\bibnamefont
			{Ye}}, \bibinfo {author} {\bibfnamefont {H.~J.}\ \bibnamefont {Kimble}}, \
		and\ \bibinfo {author} {\bibfnamefont {H.}~\bibnamefont {Katori}},\
	}\bibfield  {title} {\enquote {\bibinfo {title} {{Quantum State Engineering
					and Precision Metrology Using State-Insensitive Light Traps}},}\ }\href
	{\doibase 10.1126/science.1148259} {\bibfield  {journal} {\bibinfo  {journal}
			{Science}\ }\textbf {\bibinfo {volume} {320}},\ \bibinfo {pages} {1734}
		(\bibinfo {year} {2008})}\BibitemShut {NoStop}%
	\bibitem [{\citenamefont {Englert}\ \emph {et~al.}(2011)\citenamefont
		{Englert}, \citenamefont {Mielenz}, \citenamefont {Sommer}, \citenamefont
		{Bayerl}, \citenamefont {Motsch}, \citenamefont {Pinkse}, \citenamefont
		{Rempe},\ and\ \citenamefont {Zeppenfeld}}]{Englert2011}%
	\BibitemOpen
	\bibfield  {author} {\bibinfo {author} {\bibfnamefont {B.~G.~U.}\
			\bibnamefont {Englert}}, \bibinfo {author} {\bibfnamefont {M.}~\bibnamefont
			{Mielenz}}, \bibinfo {author} {\bibfnamefont {C.}~\bibnamefont {Sommer}},
		\bibinfo {author} {\bibfnamefont {J.}~\bibnamefont {Bayerl}}, \bibinfo
		{author} {\bibfnamefont {M.}~\bibnamefont {Motsch}}, \bibinfo {author}
		{\bibfnamefont {P.}~\bibnamefont {Pinkse}}, \bibinfo {author} {\bibfnamefont
			{G.}~\bibnamefont {Rempe}}, \ and\ \bibinfo {author} {\bibfnamefont
			{M.}~\bibnamefont {Zeppenfeld}},\ }\bibfield  {title} {\enquote {\bibinfo
			{title} {{Storage and Adiabatic Cooling of Polar Molecules in a
					Microstructured Trap}},}\ }\href {\doibase 10.1103/PhysRevLett.107.263003}
	{\bibfield  {journal} {\bibinfo  {journal} {Phys. Rev. Lett.}\ }\textbf
		{\bibinfo {volume} {107}},\ \bibinfo {pages} {263003} (\bibinfo {year}
		{2011})}\BibitemShut {NoStop}%
	\bibitem [{\citenamefont {Cornet}\ and\ \citenamefont
		{Winnewisser}(1980)}]{Cornet1980}%
	\BibitemOpen
	\bibfield  {author} {\bibinfo {author} {\bibfnamefont {R.}~\bibnamefont
			{Cornet}}\ and\ \bibinfo {author} {\bibfnamefont {G.}~\bibnamefont
			{Winnewisser}},\ }\bibfield  {title} {\enquote {\bibinfo {title} {{A precise
					study of the rotational spectrum of formaldehyde H$_2^{12}$C$^{16}$O,
					H$_2^{13}$C$^{16}$O, H$_2^{12}$C$^{18}$O, H$_2^{13}$C$^{18}$O}},}\ }\href
	{\doibase 10.1016/0022-2852(80)90154-X} {\bibfield  {journal} {\bibinfo
			{journal} {J. Mol. Spectrosc.}\ }\textbf {\bibinfo {volume} {80}},\ \bibinfo
		{pages} {438} (\bibinfo {year} {1980})}\BibitemShut {NoStop}%
	\bibitem [{\citenamefont {Zeppenfeld}(2013)}]{Zeppenfeld2013}%
	\BibitemOpen
	\bibfield  {author} {\bibinfo {author} {\bibfnamefont {M.}~\bibnamefont
			{Zeppenfeld}},\ }\emph {\bibinfo {title} {{Electric Trapping and Cooling of
				Polyatomic Molecules}}},\ \href
	{http://nbn-resolving.de/urn/resolver.pl?urn:nbn:de:bvb:91-diss-20131121-1172964-0-5}
	{\bibinfo {type} {Dissertation}},\ \bibinfo  {school} {Technische
		Universit{\"{a}}t M{\"{u}}nchen} (\bibinfo {year} {2013})\BibitemShut
	{NoStop}%
	\bibitem [{\citenamefont {Gl{\"{o}}ckner}\ \emph
		{et~al.}(2015{\natexlab{a}})\citenamefont {Gl{\"{o}}ckner}, \citenamefont
		{Prehn}, \citenamefont {Rempe},\ and\ \citenamefont
		{Zeppenfeld}}]{Gloeckner2015}%
	\BibitemOpen
	\bibfield  {author} {\bibinfo {author} {\bibfnamefont {R.}~\bibnamefont
			{Gl{\"{o}}ckner}}, \bibinfo {author} {\bibfnamefont {A.}~\bibnamefont
			{Prehn}}, \bibinfo {author} {\bibfnamefont {G.}~\bibnamefont {Rempe}}, \ and\
		\bibinfo {author} {\bibfnamefont {M.}~\bibnamefont {Zeppenfeld}},\ }\bibfield
	{title} {\enquote {\bibinfo {title} {{Rotational state detection of
					electrically trapped polyatomic molecules}},}\ }\href {\doibase
		10.1088/1367-2630/17/5/055022} {\bibfield  {journal} {\bibinfo  {journal}
			{New J. Phys.}\ }\textbf {\bibinfo {volume} {17}},\ \bibinfo {pages} {055022}
		(\bibinfo {year} {2015}{\natexlab{a}})}\BibitemShut {NoStop}%
	\bibitem [{\citenamefont {Townes}\ and\ \citenamefont
		{Schawlow}(1975)}]{Townes1975}%
	\BibitemOpen
	\bibfield  {author} {\bibinfo {author} {\bibfnamefont {C.~H.}\ \bibnamefont
			{Townes}}\ and\ \bibinfo {author} {\bibfnamefont {A.~L.}\ \bibnamefont
			{Schawlow}},\ }\href@noop {} {\emph {\bibinfo {title} {{Microwave
					Spectroscopy}}}}\ (\bibinfo  {publisher} {Dover},\ \bibinfo {address} {New
		York},\ \bibinfo {year} {1975})\BibitemShut {NoStop}%
	\bibitem [{\citenamefont {Fabricant}\ \emph {et~al.}(1977)\citenamefont
		{Fabricant}, \citenamefont {Krieger},\ and\ \citenamefont
		{Muenter}}]{Fabricant1977}%
	\BibitemOpen
	\bibfield  {author} {\bibinfo {author} {\bibfnamefont {B.}~\bibnamefont
			{Fabricant}}, \bibinfo {author} {\bibfnamefont {D.}~\bibnamefont {Krieger}},
		\ and\ \bibinfo {author} {\bibfnamefont {J.~S.}\ \bibnamefont {Muenter}},\
	}\bibfield  {title} {\enquote {\bibinfo {title} {{Molecular beam electric
					resonance study of formaldehyde, thioformaldehyde, and ketene}},}\ }\href
	{\doibase 10.1063/1.434988} {\bibfield  {journal} {\bibinfo  {journal} {J.
				Chem. Phys.}\ }\textbf {\bibinfo {volume} {67}},\ \bibinfo {pages} {1576}
		(\bibinfo {year} {1977})}\BibitemShut {NoStop}%
	\bibitem [{\citenamefont {Br{\"{u}}nken}\ \emph {et~al.}(2003)\citenamefont
		{Br{\"{u}}nken}, \citenamefont {M{\"{u}}ller}, \citenamefont {Lewen},\ and\
		\citenamefont {Winnewisser}}]{Brunken2003}%
	\BibitemOpen
	\bibfield  {author} {\bibinfo {author} {\bibfnamefont {S.}~\bibnamefont
			{Br{\"{u}}nken}}, \bibinfo {author} {\bibfnamefont {H.~S.~P.}\ \bibnamefont
			{M{\"{u}}ller}}, \bibinfo {author} {\bibfnamefont {F.}~\bibnamefont {Lewen}},
		\ and\ \bibinfo {author} {\bibfnamefont {G.}~\bibnamefont {Winnewisser}},\
	}\bibfield  {title} {\enquote {\bibinfo {title} {{High accuracy measurements
					on the ground state rotational spectrum of formaldehyde (H$_2$CO) up to 2
					THz}},}\ }\href {\doibase 10.1039/b301657a} {\bibfield  {journal} {\bibinfo
			{journal} {Phys. Chem. Chem. Phys.}\ }\textbf {\bibinfo {volume} {5}},\
		\bibinfo {pages} {1515} (\bibinfo {year} {2003})}\BibitemShut {NoStop}%
	\bibitem [{\citenamefont {Zeppenfeld}\ \emph {et~al.}(2009)\citenamefont
		{Zeppenfeld}, \citenamefont {Motsch}, \citenamefont {Pinkse},\ and\
		\citenamefont {Rempe}}]{Zeppenfeld2009}%
	\BibitemOpen
	\bibfield  {author} {\bibinfo {author} {\bibfnamefont {M.}~\bibnamefont
			{Zeppenfeld}}, \bibinfo {author} {\bibfnamefont {M.}~\bibnamefont {Motsch}},
		\bibinfo {author} {\bibfnamefont {P.}~\bibnamefont {Pinkse}}, \ and\ \bibinfo
		{author} {\bibfnamefont {G.}~\bibnamefont {Rempe}},\ }\bibfield  {title}
	{\enquote {\bibinfo {title} {{Optoelectrical cooling of polar molecules}},}\
	}\href {\doibase 10.1103/PhysRevA.80.041401} {\bibfield  {journal} {\bibinfo
			{journal} {Phys. Rev. A}\ }\textbf {\bibinfo {volume} {80}},\ \bibinfo
		{pages} {041401(R)} (\bibinfo {year} {2009})}\BibitemShut {NoStop}%
	\bibitem [{\citenamefont {Zeppenfeld}\ \emph {et~al.}(2012)\citenamefont
		{Zeppenfeld}, \citenamefont {Englert}, \citenamefont {Gl{\"{o}}ckner},
		\citenamefont {Prehn}, \citenamefont {Mielenz}, \citenamefont {Sommer},
		\citenamefont {van Buuren}, \citenamefont {Motsch},\ and\ \citenamefont
		{Rempe}}]{Zeppenfeld2012}%
	\BibitemOpen
	\bibfield  {author} {\bibinfo {author} {\bibfnamefont {M.}~\bibnamefont
			{Zeppenfeld}}, \bibinfo {author} {\bibfnamefont {B.~G.~U.}\ \bibnamefont
			{Englert}}, \bibinfo {author} {\bibfnamefont {R.}~\bibnamefont
			{Gl{\"{o}}ckner}}, \bibinfo {author} {\bibfnamefont {A.}~\bibnamefont
			{Prehn}}, \bibinfo {author} {\bibfnamefont {M.}~\bibnamefont {Mielenz}},
		\bibinfo {author} {\bibfnamefont {C.}~\bibnamefont {Sommer}}, \bibinfo
		{author} {\bibfnamefont {L.~D.}\ \bibnamefont {van Buuren}}, \bibinfo
		{author} {\bibfnamefont {M.}~\bibnamefont {Motsch}}, \ and\ \bibinfo {author}
		{\bibfnamefont {G.}~\bibnamefont {Rempe}},\ }\bibfield  {title} {\enquote
		{\bibinfo {title} {{Sisyphus cooling of electrically trapped polyatomic
					molecules.}}}\ }\href {\doibase 10.1038/nature11595} {\bibfield  {journal}
		{\bibinfo  {journal} {Nature}\ }\textbf {\bibinfo {volume} {491}},\ \bibinfo
		{pages} {570} (\bibinfo {year} {2012})}\BibitemShut {NoStop}%
	\bibitem [{\citenamefont {Gl{\"{o}}ckner}\ \emph
		{et~al.}(2015{\natexlab{b}})\citenamefont {Gl{\"{o}}ckner}, \citenamefont
		{Prehn}, \citenamefont {Englert}, \citenamefont {Rempe},\ and\ \citenamefont
		{Zeppenfeld}}]{Glockner2015a}%
	\BibitemOpen
	\bibfield  {author} {\bibinfo {author} {\bibfnamefont {R.}~\bibnamefont
			{Gl{\"{o}}ckner}}, \bibinfo {author} {\bibfnamefont {A.}~\bibnamefont
			{Prehn}}, \bibinfo {author} {\bibfnamefont {B.~G.~U.}\ \bibnamefont
			{Englert}}, \bibinfo {author} {\bibfnamefont {G.}~\bibnamefont {Rempe}}, \
		and\ \bibinfo {author} {\bibfnamefont {M.}~\bibnamefont {Zeppenfeld}},\
	}\bibfield  {title} {\enquote {\bibinfo {title} {{Rotational Cooling of
					Trapped Polyatomic Molecules}},}\ }\href {\doibase
		10.1103/PhysRevLett.115.233001} {\bibfield  {journal} {\bibinfo  {journal}
			{Phys. Rev. Lett.}\ }\textbf {\bibinfo {volume} {115}},\ \bibinfo {pages}
		{233001} (\bibinfo {year} {2015}{\natexlab{b}})}\BibitemShut {NoStop}%
	\bibitem [{Note1()}]{Note1}%
	\BibitemOpen
	\bibinfo {note} {With the proof-of-concept demonstration of our method, the
		uncertainty of the zero-field center frequency could be reduced by a factor
		of $\sim $\protect \tmspace +\thinmuskip {.1667em}10 compared to the most
		precise direct measurement reported in the literature~\cite
		{Cornet1980}.}\BibitemShut {Stop}%
\end{thebibliography}
\end{document}